\documentclass[11pt]{article}
\usepackage{geometry}                
\usepackage{graphicx}
\usepackage{amssymb}
\usepackage{epstopdf}
\title{The Sensitivity of SNO+ to $\Delta m_{12}^{2}$ Using Reactor Anti-neutrino Data}
\author{Eugene Guillian\footnote{Contact: guillian@owl.phy.queensu.ca or ehguillian@gmail.com}}

\begin{document}
\maketitle

\abstract{Insofar as the detection of anti-neutrinos from nuclear reactors is concerned, the SNO+ detector -- a 1 kilo-tonne liquid scintillator detector that inherits the experimental infrastructure from the recently finished SNO experiment -- is expected to perform just as well as the KamLAND experiment.  The most important difference between these experiments is the distribution of nuclear reactors: whereas KamLAND has 9 nuclear reactor sites within 300~km with a flux-averaged baseline of about 180~km, SNO+ has only 1 within 300~km, with an average baseline of $\approx 750$~km.  As a result, the reactor anti-neutrino flux at SNO+ is only about 1/5 that at KamLAND, and the ability of SNO+ to constrain the solar neutrino oscillation parameter is diminished by a factor of about $\sqrt{1/5} = 1/2.2$ relative to KamLAND.  In spite of this, SNO+ has comparable sensitivity to $\Delta m^{2}_{12}$ as KamLAND because the rate of change of the spectral distortion as a function of this parameter is much greater than for KamLAND.  In this report, this advantage is examined quantitatively using a geometric approximation that makes clear how the shape from SNO+ has more statistical power than that from KamLAND.  This result then is confirmed by determining the sensitivity to $\Delta m^{2}_{12}$ using an ensemble experiment technique.}

\section{Introduction}

\begin{figure}[h] 
\begin{center}
   \includegraphics[width=\textwidth]{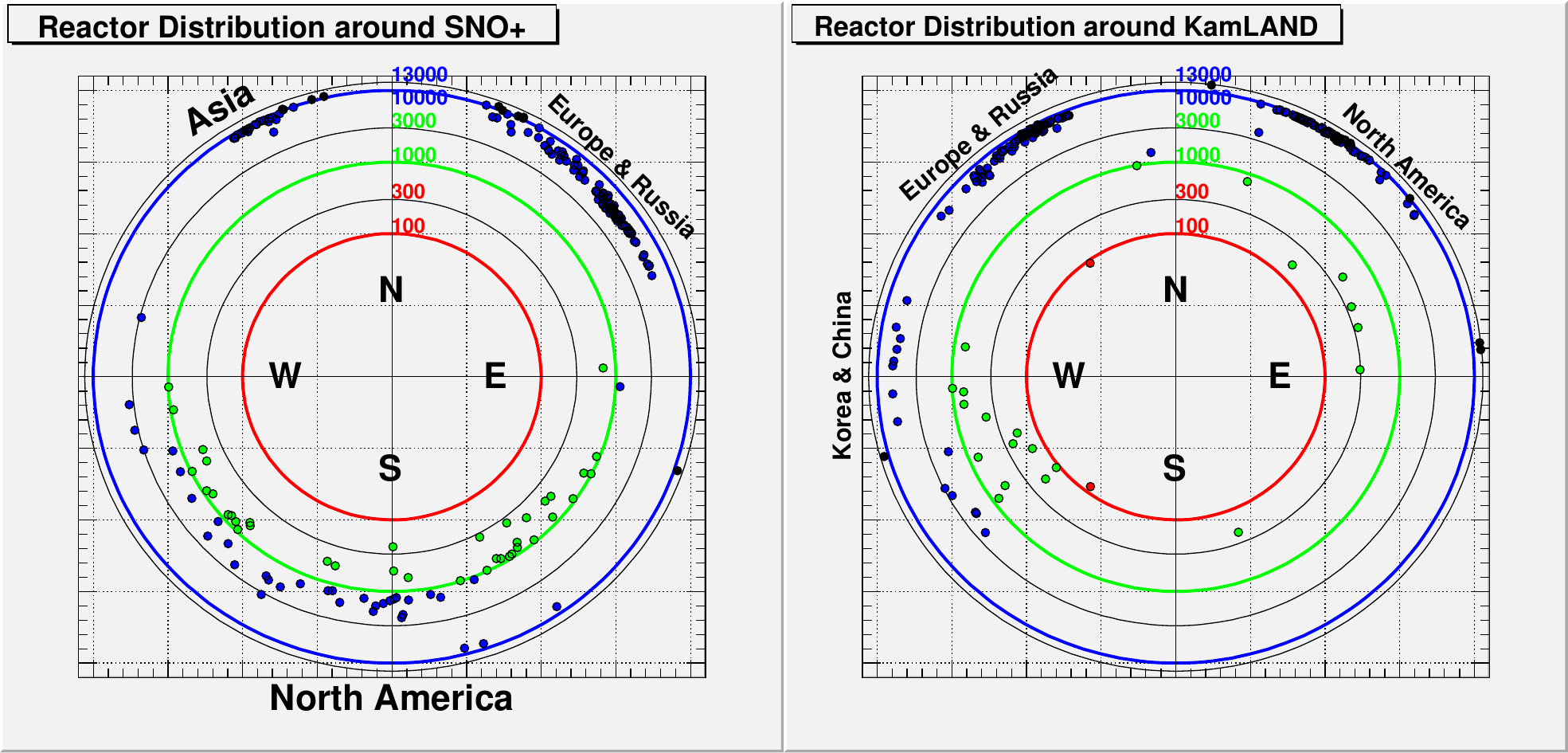} 
\end{center}
   \caption{The distribution of nuclear reactor sites in horizontal coordinates centered at SNO+ (left) and at KamLAND (right).  The center represents the detector position, while each point represents a nuclear reactor site, which may contain multiple reactors.  The radius from the center represents the distance from the reactor, in logarithmic scale.  The concentric circles show distances of 100~km, 300~km, and so on out to 13,000~km.  KamLAND has two sites within 100~km and nine within 300~km, while SNO+ has none within 100~km and just one within 300~km.}
   \label{fig:reactor_distribution}
\end{figure}

The SNO+ detector is a 1-kton liquid scintillator detector that can detect reactor anti-neutrinos via inverse beta decay.  The performance of SNO+ in the detection of these anti-neutrinos is basically identical to that of KamLAND: the target masses are almost identical, the detection efficiency and detector live time fraction are almost the same, and the background noise levels are similar.  The main difference between the two experiments is the distribution of the sources of anti-neutrinos (Fig.~\ref{fig:reactor_distribution}).  Kamioka is surrounded by two reactor sites less than 100~km away and nine within 300~km, with a flux-weighted average of about 180~km; in contrast, SNO+ has no reactor sites within a distance of 100~km, and just one within 300~km, with a corresponding flux-weighted average distance of $\approx$~750~km.  As a result of these differences, the reactor anti-neutrino flux at SNO+ is about 1/5 of that at KamLAND.

In spite of the lower flux at SNO+, its sensitivity to the neutrino oscillation parameter $\Delta m_{12}^{2}$ for a given exposure (say, in units of $10^{32}$~proton-years) is almost the same as for KamLAND.  The statistical disadvantage of SNO+ is compensated by the more distinct oscillatory signature of the anti-neutrino energy spectrum.  In general, for a nearly background-free measurement above $E_{\nu} = 3.4$~MeV (this energy cut is applied to avoid geo-neutrinos, accidentals, and the $^{13}$C($\alpha$,n)$^{16}$O background), the sensitivity to $\Delta m_{12}^{2}$ depends on the total number of detected events $N$ and a factor $K$ that depends on the rate of change of the shape of the distortion in the energy spectrum as a function of $\Delta m_{12}^{2}$.  The fractional uncertainty of $\Delta m_{12}^{2}$ is proportional to the following:

\begin{equation}
\frac{\delta\left(\Delta m^{2}_{12}\right)}{\Delta m^{2}_{12}} \propto \frac{K}{\sqrt{N}}
\end{equation}

\noindent The statistical disadvantage of SNO+ compared to KamLAND is expressed as \\ $\sqrt{N_{KamLAND}/N_{SNO+}} \approx \sqrt{5} \approx 2.2$.  The purpose of this document is to show that $K_{KamLAND}/K_{SNO+}$ has approximately the same value, so that the sensitivity of these two experiments to $\Delta m_{12}^{2}$ is almost the same.

\section{Basic Information Used in This Study}

The latest result from KamLAND~\cite{kamland} is used as a baseline for the comparison with SNO+.  Specifically, the exposure is chosen to be $2.44 \times 10^{32}$~proton-years, and the best-fit solar neutrino oscillation parameter values was chosen as:

\begin{eqnarray}
\label{eqn:dmsq_best_fit}
\Delta m_{12}^{2} & = & 7.55\times10^{-5} \; \mbox{eV}^{2} \\
\label{eqn:ss2t_best_fit}
\sin^{2} 2\theta_{\odot} & = & 0.928
\end{eqnarray}

\noindent The mixing angle expressed in terms of $\tan^{2}\theta_{\odot}$ is:

\begin{equation}
\tan^{2}\theta_{\odot} = 0.577
\end{equation}

\noindent These values are not exactly the same as given in~\cite{kamland}, but they are well within the uncertainties.

The choice of uncertainty to use is not unique, since several different analyses are presented in~\cite{kamland}.  Since the aim of this study is to compare the statistical power of SNO+ with that of KamLAND, the errors for the analysis using only the KamLAND data are used (in preference to that from the global analysis of data from KamLAND and all solar neutrino experiments).  Combining the statistical and systematic errors in quadrature, the 1$\sigma$ uncertainty on the oscillation parameters are as follows:

\begin{equation}
\Delta m_{12}^{2} = (7.55^{+0.21}_{-0.20}) \times 10^{-5} \; \mbox{eV}^{2}
\end{equation}

\begin{equation}
\tan^{2}\theta_{\odot} = 0.577^{+0.140}_{-0.090}
\end{equation}

\noindent The 1$\sigma$ range of $\sin^{2}2\theta_{\odot}$ based on the above is $\left[0.881, 0.973\right]$.  These ranges were used as a guide in determining the parameter range to examine in this study.

The location and power of nuclear reactors was obtained using the information provided by the International Nuclear Safety Center (INSC) of the Argonne National Laboratory (ANL)~\cite{insc}.  The information was obtained in 2006; for details about the list used in this study, see Appendix~\ref{app:reactors}.  It is less accurate than information provided by the reactor operators, such as that used by the KamLAND collaboration.  In particular, absolute normalization of the anti-neutrino flux is rather inaccurate because the rated power is often different from the power during operation, and because the INSC list has no information about the duty cycle.  In this study, we used a constant factor of 0.8 to represent the average duty cycle for all reactors ({\it i.e.} the power of all reactors was scaled down by this factor).

The anti-neutrino detection rate depends not only on the flux, but on the detection efficiency.  This was tuned in our model so that the number of anti-neutrino events from our model matches that from KamLAND.  Although KamLAND's detection efficiency is energy dependent, we ignored this dependence.  As discussed below, a detection efficiency of 95\% for $E_{\nu} > 3.4$~MeV was found to give a good match between the model and KamLAND.

The above inaccuracy of the nuclear reactor model also affects the predicted shape of the anti-neutrino spectrum after oscillations, but to a lesser extent that the overall normalization of the flux.  Below, we shall show that the spectrum shape from our model is quite similar to that from KamLAND.  Since the goal of this study is to compare the sensitivity of the two experiments to $\Delta m^{2}_{12}$, the approximation should be adequate as long as the basic features in the spectrum shape are reproduced.

\section{Tuning the Nuclear Reactor Model to KamLAND Data}

\begin{figure}[h] 
\begin{center}
   \includegraphics[width=12cm]{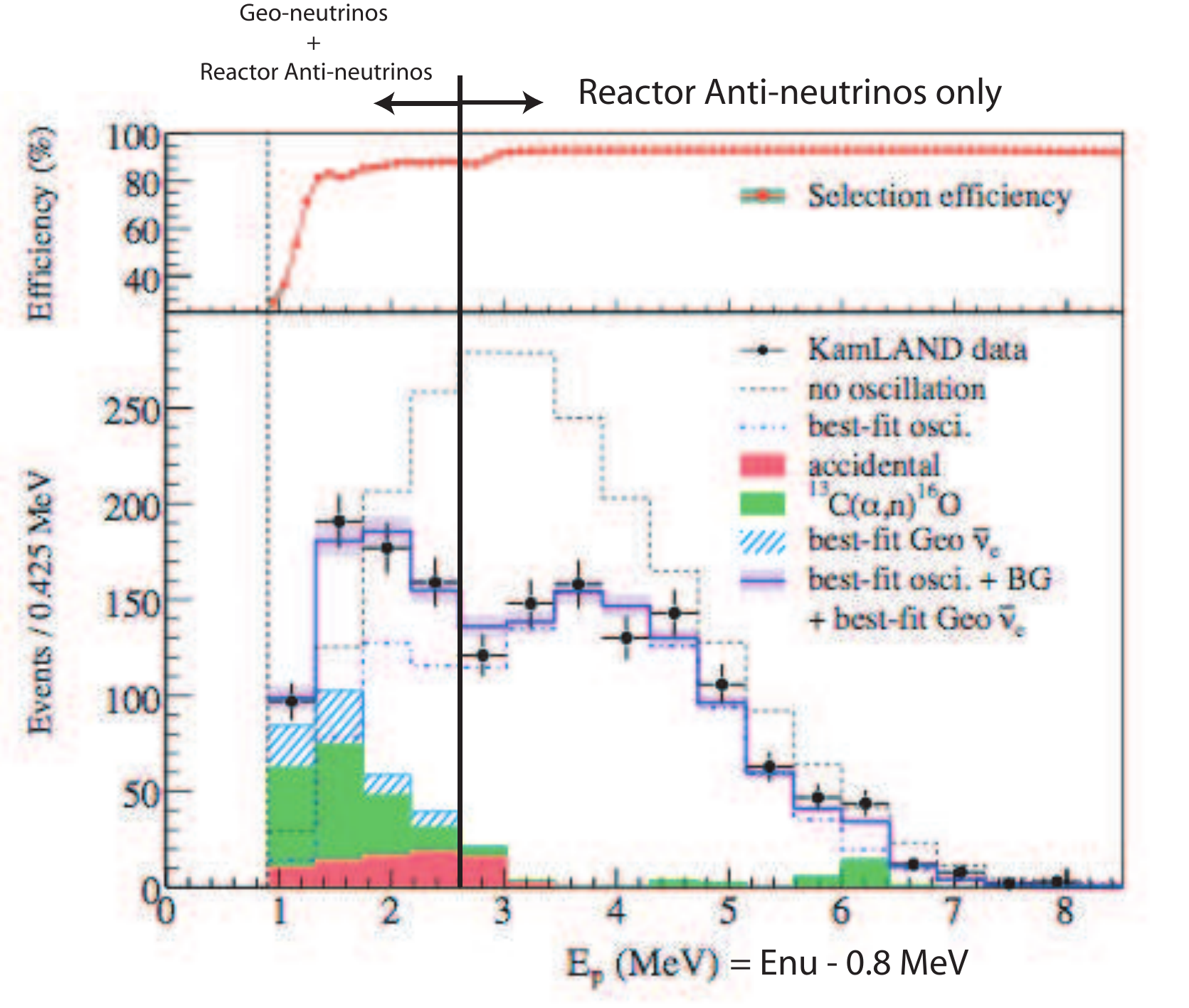} 
\end{center}
   \caption{KamLAND's reactor anti-neutrino spectrum, adopted from Fig.1 of~\cite{kamland}.  The horizontal axis indicates prompt energy, which is $E_{\nu} - 0.8$~MeV.  The vertical line at $E_{p} = 2.6$~MeV ($E_{\nu} = 3.4$~MeV) divides the data into an almost background-free reactor anti-neutrino region above this energy and the region below which is contaminated by various sources of background noise.}
   \label{fig:kamland_nuebar_spec}
\end{figure}

According to the latest KamLAND result~\cite{kamland}, the expected number of events with $E_{\nu} > 1.8$~MeV passing all selection cuts before oscillations is 2179.  In comparison, the reactor model used here predicts 3345 events, assuming 100\% duty cycle and detection efficiency.  Taking the duty cycle to be 80\%, this becomes 2676 events.  Taking the ratio of 2179 to 2676, one has a tuned detection efficiency for $E_{\nu} > 1.8$~MeV of 0.81.  This is a reasonable value in view of KamLAND's detection efficiency shown in the top frame of Fig.~\ref{fig:kamland_nuebar_spec}.

\begin{figure}[h] 
\begin{center}
   \includegraphics[width=12cm]{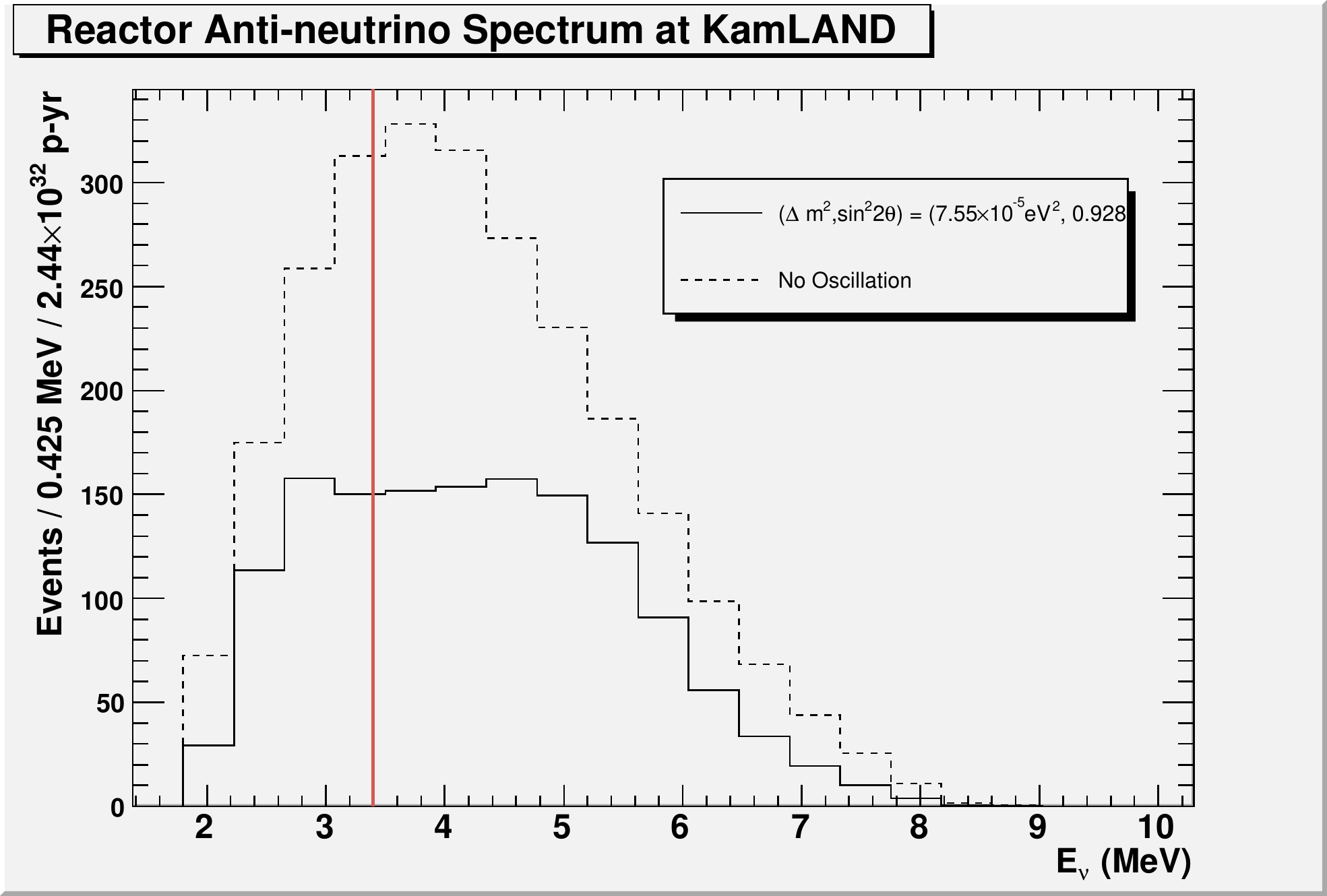} 
\end{center}
   \caption{The model reactor anti-neutrino spectrum before and after neutrino oscillations.  The oscillated spectrum used the best-fit solar neutrino oscillation parameters in Eqs.~\ref{eqn:dmsq_best_fit}-\ref{eqn:ss2t_best_fit}.}
   \label{fig:reactor_neutrino_spec_kl}
\end{figure}

In the study presented here, we shall focus on the essentially background-free region of $E_{\nu} > 3.4$~MeV, where KamLAND's detection efficiency is mostly well above 90\%.  In order to tune our model, we shall use as a standard the number of events after oscillations in the bin $E_{p} \in \left[3.6, 4.0\right]$~MeV ($E_{\nu} \in \left[4.4,4.8\right]$)~MeV, which is about 160.  Fig.~\ref{fig:reactor_neutrino_spec_kl} shows the model spectrum where the number of events after oscillations in the reference bin was tuned to be approximately the same as for KamLAND.  The detection efficiency in the model after tuning was 95\%.  The agreement between the model and KamLAND spectra is reasonable.  There is some disagreement around $E_{\nu} = 3.4$~MeV because the energy dependence of the detection efficiency has been ignored in the model.  But this level of disagreement should not significantly affect conclusions regarding the relative sensitivity to $\Delta m^{2}_{12}$ in KamLAND and SNO+.

\section{The Shape of the Reactor Anti-neutrino Spectrum}

\begin{figure}[h] 
\begin{center}
   \includegraphics[width=\textwidth]{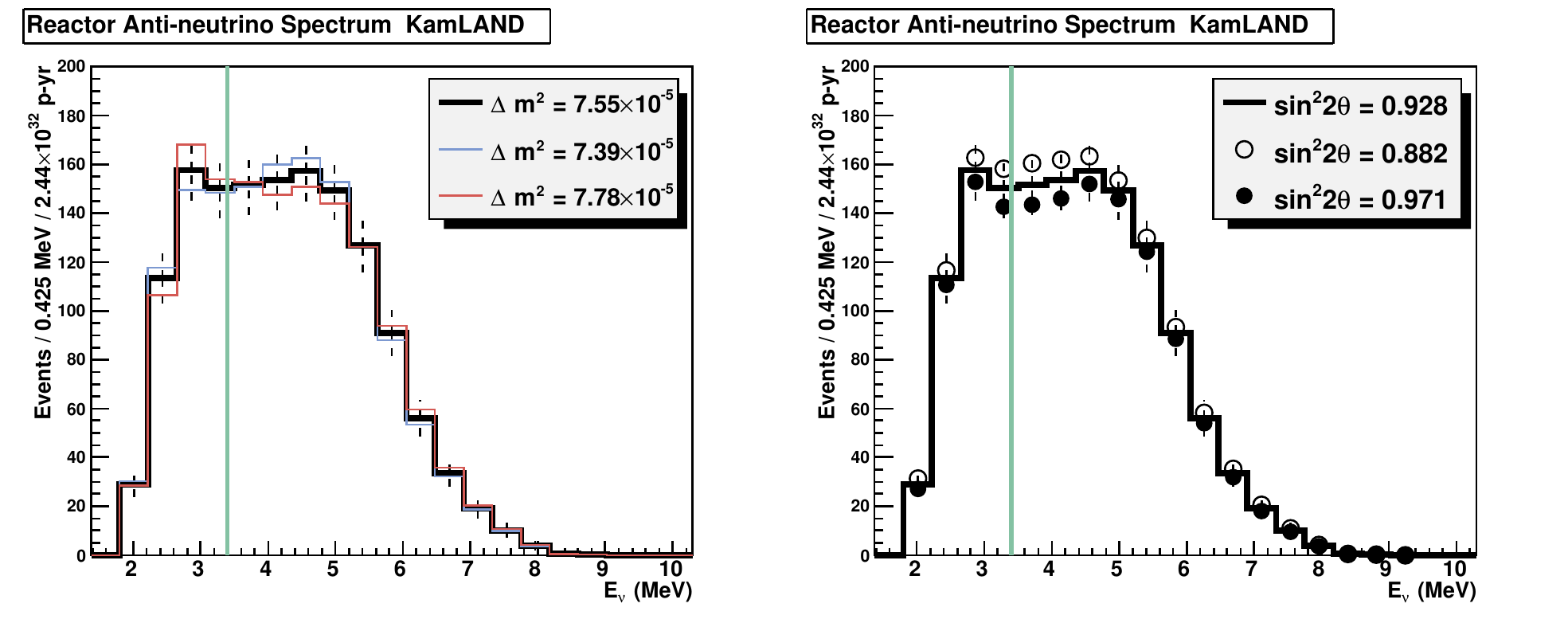} 
\end{center}
   \caption{The change in spectrum as a function of neutrino oscillation parameter values in KamLAND.  In both frames, the histogram in thick black is for $(\Delta m^{2}, \sin^{2}2\theta) = (7.55\times10^{-5}~\mbox{eV}^{2}, 0.928)$.  In the left frame, the blue and red histograms show the change in the spectrum when $\Delta m^{2}$ is changed by 1$\sigma$, while the mixing angle is kept fixed.  In the right frame, the mixing angle is changed by 1$\sigma$ while $\Delta m^{2}$ is kept fixed.  The error bar shows the statistical error in each bin for the thick black histogram.}
   \label{fig:reactor_neutrino_spec_diff_par_vals_v2_kl}
\end{figure}

\begin{figure}[h] 
\begin{center}
   \includegraphics[width=\textwidth]{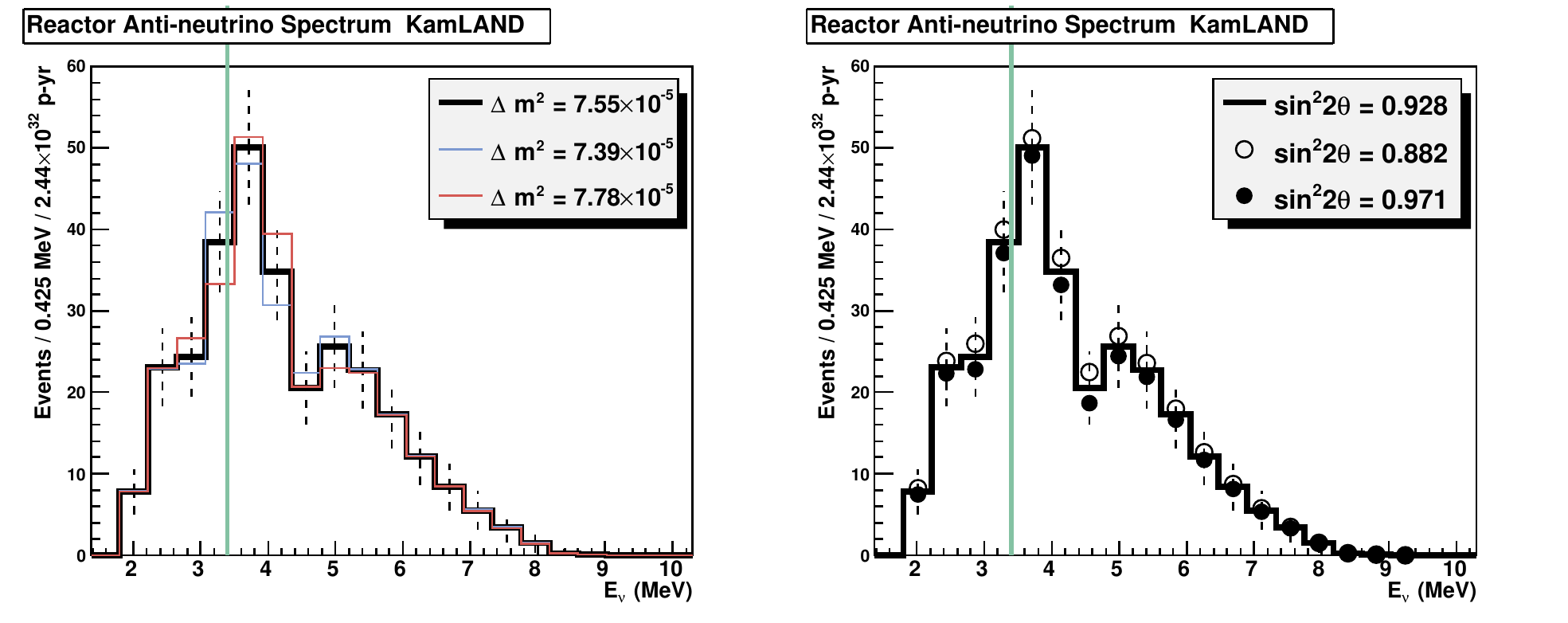} 
\end{center}
   \caption{Same as Fig.~\ref{fig:reactor_neutrino_spec_diff_par_vals_v2_kl}, but for SNO+.}
   \label{fig:reactor_neutrino_spec_diff_par_vals_v2_sno+}
\end{figure}

The sensitivity of a given experiment to the neutrino oscillation parameters depends on how quickly the event rate and the spectrum shape change as a function of the parameter values.  This is illustrated for KamLAND and SNO+ in Figs.~\ref{fig:reactor_neutrino_spec_diff_par_vals_v2_kl} and \ref{fig:reactor_neutrino_spec_diff_par_vals_v2_sno+}.  There are a few notable features about the two figures.

\begin{itemize}
\item{The event statistics in KamLAND is about 5 times as large as that in SNO+.}
\item{In KamLAND, when $\Delta m^{2}$ is changed by 1$\sigma$, the histogram contents in many bins changes by somewhat less than 1$\sigma$.  The same holds when $\sin^{2}2\theta$ is changed by 1$\sigma$.}
\item{In SNO+, when $\Delta m^{2}$ is changed by 1$\sigma$, the {\it relative} change in shape of the spectrum in many bins is much larger than is the case for KamLAND.  On the other hand, the change in the spectrum with a 1$\sigma$ change in $\sin^{2}2\theta$ is much smaller than in KamLAND.}
\end{itemize}

\noindent The above indicate that, because the event rate is much smaller in SNO+ than in KamLAND, SNO+ is much less sensitive to the mixing angle.  However, because the rate of change of spectrum shape as a function of $\Delta m^{2}$ in SNO+ is much larger than in KamLAND, SNO+, for a given number of events, is more sensitive to $\Delta m^{2}$.  Of course, this superiority in sensitivity is, to some extent, canceled out by the fact that the event rate in SNO+ is only 1/5 that in KamLAND. 

The advantage that SNO+ has over KamLAND in its sensitivity to $\Delta m^{2}$ can be expressed quantitatively using a quantity $K$, which is related to the fractional error in $\Delta m^{2}$ as follows:

\begin{equation}
\label{eqn:delta_dmsq_over_dmsq}
\frac{\delta \left( \Delta m^{2}\right)}{\Delta m^{2}} = \alpha \cdot \frac{K}{\sqrt{N}} \; ,
\end{equation}

\noindent where $N$ is the number of observed events, $K$ is a constant that depends on how quickly the shape of the anti-neutrino energy spectrum changes as a function of $\Delta m^{2}$, and $\alpha$ is a constant that is the same for any experiment (it depends on how the likelihood function is defined).  The quantity $K$ is given by the following expression:

\begin{equation}
\frac{1}{K^{2}} = \int \; dE_{\nu} \; \frac{\left[ f(E_{\nu}; \Delta m^{2}_{1}) - f(E_{\nu}; \Delta m^{2}_{0}) \right]^{2}}{f(E_{\nu}; \Delta m^{2}_{0})}
\end{equation}

\noindent In the above integral, the function $f(E_{\nu}; \Delta m^{2})$ is the reactor antineutrino energy spectrum for $\Delta m^{2}$ (the mixing angle is assumed to be fixed such that $\sin^{2}2\theta = 0.928$).  It is normalized so that the integral is equal to unity:

\begin{equation}
\int \; dE_{\nu} \; f(E_{\nu}; \Delta m^{2}) = 1
\end{equation}

\begin{figure}[h] 
\begin{center}
   \includegraphics[width=12cm]{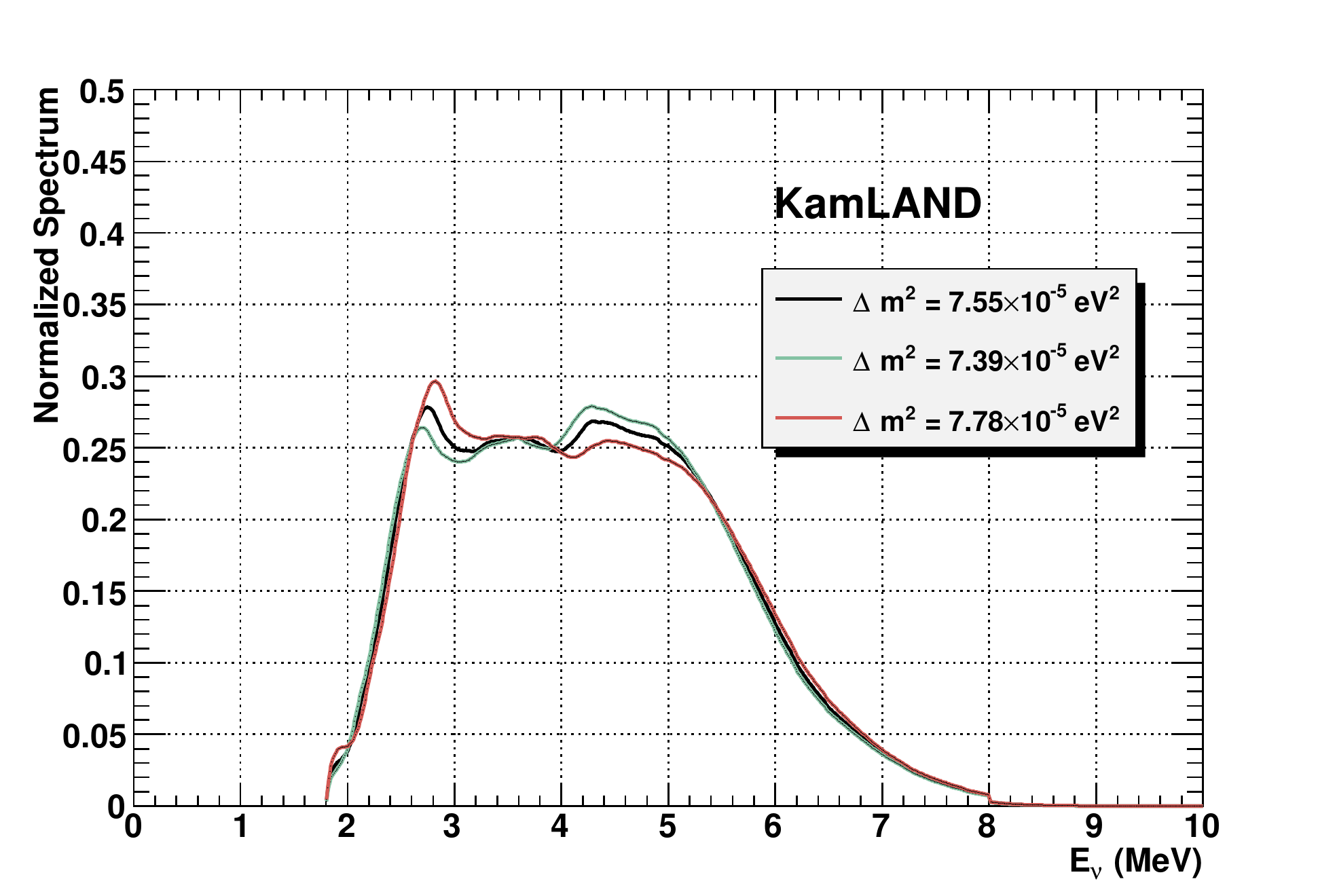} 
\end{center}
   \caption{The normalized reactor anti-neutrino energy spectrum for KamLAND for $\Delta m^{2}$ at the central value (black) and at $\pm 1\sigma$.  The mixing angle is such that $\sin^{2}2\theta = 0.928$.}
   \label{fig:k_fac_plots_kl}
\end{figure}

\begin{figure}[h] 
\begin{center}
   \includegraphics[width=12cm]{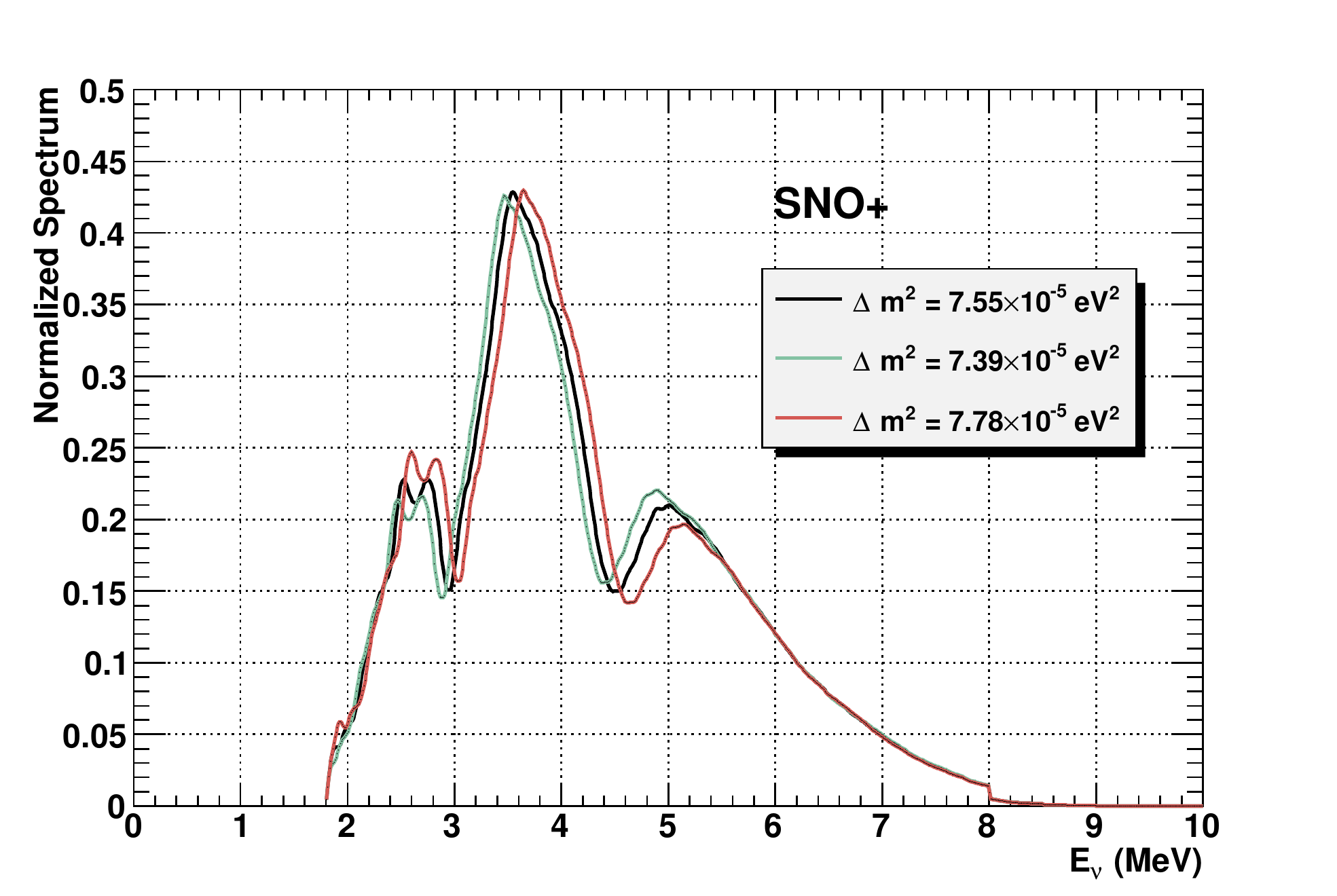} 
\end{center}
   \caption{Same as Fig.~\ref{fig:k_fac_plots_kl}, but for SNO+.}
   \label{fig:k_fac_plots_sno+}
\end{figure}

\noindent The integral for $1/K^{2}$ has a form resembling $\chi^{2}$, which is not an accident because it is extracted from a log-likelihood function for the fit parameter $\Delta m^{2}$.  In applying this formula, one takes the red or blue function in Fig.~\ref{fig:k_fac_plots_kl} or \ref{fig:k_fac_plots_sno+} and the black function and calculate the difference squared.  This is divided by the contents of the black function.  The result is the integrand, which is integrated over the energy range of the fit.  The more different the two functions, the larger the integral, which implies that $K$ is smaller.  Since $K$ is proportional to the parameter uncertainty, one can conclude that the greater the difference in shape for a unit change in parameter value, the smaller the parameter error.

The value of $K$ for KamLAND and SNO+ is shown in Table~\ref{tab:k_factors}.  The quantity $K_{-}$ is the $K$-factor for the case where the function at the central value of $\Delta m^{2}$ is compared to that at $-1\sigma$, while $K_{+}$ is that for $+1\sigma$.  The final row in the table shows the ratio of the factors, KamLAND to SNO+.  The factor at KamLAND is about 2.6 to 2.8 times larger than at SNO+.  This is a quantitative measure of how much more power SNO+ has compared to KamLAND in extracting $\Delta m^{2}$ due to the rate of change of the shape as a function of $\Delta m^{2}$.  Specifically, the ratio of the fractional error of $\Delta m^{2}$ between KamLAND and SNO+ is (from Eqn.~\ref{eqn:delta_dmsq_over_dmsq}):

\begin{table}
\begin{center}
\begin{tabular}{|c|c|c|} \hline\hline
 & $K_{-}$ & $K_{+}$ \\ \hline\hline
 KamLAND & 10.3 & 7.93 \\ \hline
 SNO+ & 3.97 & 2.85 \\ \hline\hline
 KamLAND/SNO+ & 2.59 & 2.78 \\ \hline\hline
\end{tabular}
\end{center}
\caption{The $K$ factors for KamLAND, SNO+, and the ratio of the factors, KamLAND to SNO+.  $K_{-}$ is the factor for the comparing the function at the central value of $\Delta m^{2}$ to that at $-1\sigma$, while $K_{+}$ is that for $+ 1\sigma$.  The integral was evaluated over $E_{\nu} \in \left[3.4, 10.0\right]$~MeV.}
\label{tab:k_factors}
\end{table}

\begin{eqnarray}
\frac{\left[ \delta \left( \Delta m^{2}\right) / (\Delta m^{2}) \right]_{\mbox{KamLAND}}}{\left[ \delta \left( \Delta m^{2}\right) / (\Delta m^{2}) \right]_{\mbox{SNO+}}} & = & \frac{\alpha \cdot \left( K/\sqrt{N} \right)_{\mbox{KamLAND}}}{\alpha \cdot \left( K/\sqrt{N} \right)_{\mbox{SNO+}}} \\
& = & \frac{K_{\mbox{KamLAND}}}{K_{\mbox{SNO+}}} \cdot \sqrt{\frac{N_{\mbox{SNO+}}}{N_{\mbox{KamLAND}}}} \\
& = & \left( 2.6 \sim 2.8 \right) \cdot \sqrt{1/5} \\ 
& = & \frac{2.6 \sim 2.8}{2.2} \\
\label{eqn:k_factor_eval}
& = & 1.2 \sim 1.3
\end{eqnarray}

\noindent This is a remarkable result: although KamLAND has five times the statistics as SNO+, the latter's sensitivity to $\Delta m^{2}$ is 20$\sim$30\% better because of the advantage in the shape of the spectrum, as indicated by the $K$-factors.  The argument based on the $K$-factor is an approximation, but the conclusion is supported by a rigorous determination of the parameter sensitivity, which is discussed in the next section.

\section{The Sensitivity of the Experiments to $\Delta m_{12}^{2}$}

The sensitivity of the experiments to $\Delta m^{2}_{12}$ was established rigorously by performing an ensemble experiment on randomly generated data.  The random data were generated using the central value of the neutrino oscillation parameters: $(\Delta m^{2}_{12}, \sin^{2}2\theta) = (7.55\times10^{-5}~\mbox{eV}^{2}, 0.928)$.  This was compared to the predicted spectrum at some other parameter pair using the following log-likelihood function:

\begin{equation}
{\cal L} = \sum_{i = 1}^{N_{bin}} \left[ \mu_{i} - n_{i} \cdot \log \mu_{i} + \log \Gamma \left( n_{i} + 1 \right) \right]
\end{equation}

\begin{figure}[h] 
\begin{center}
   \includegraphics[width=14cm]{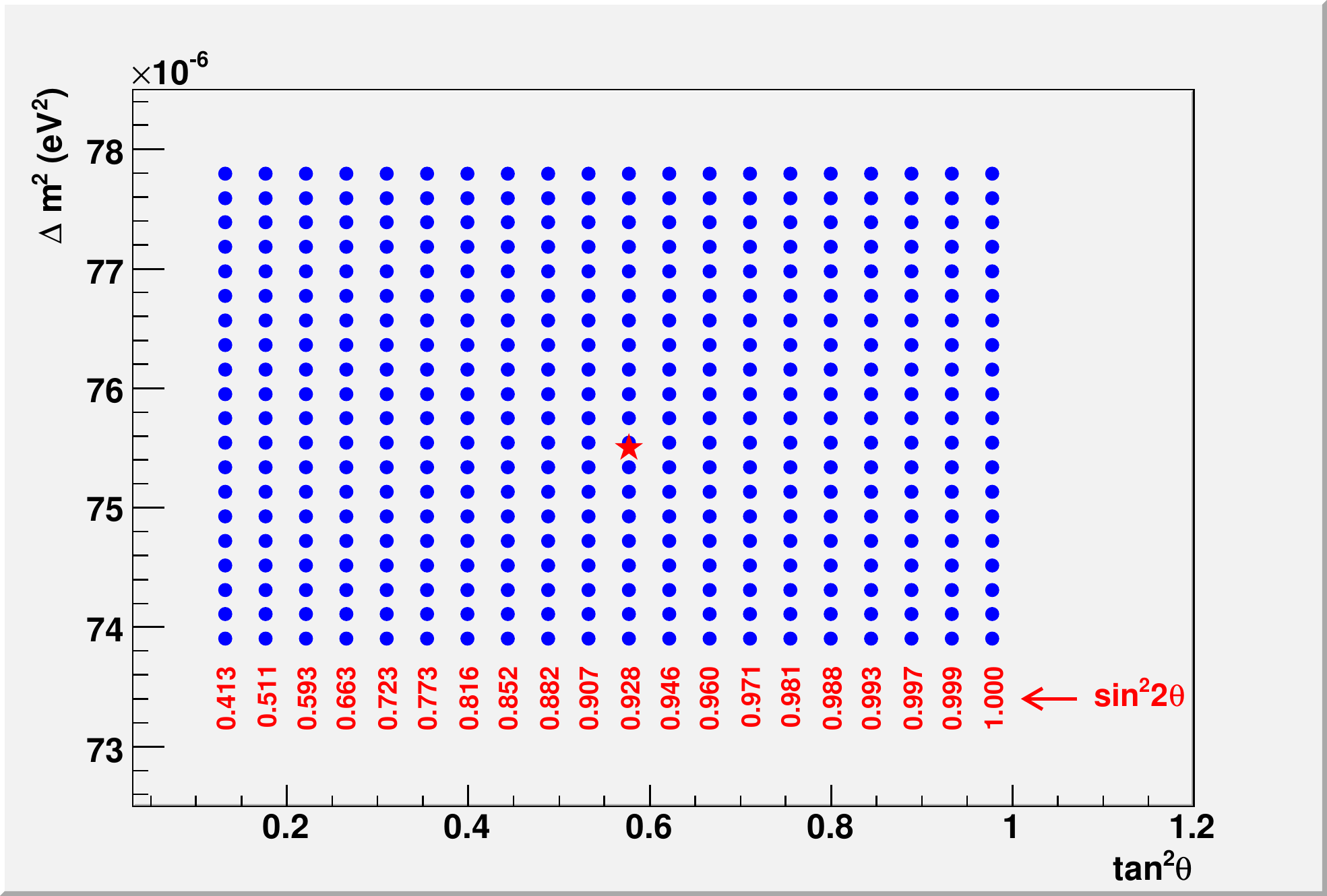} 
\end{center}
   \caption{The grid in the neutrino oscillation parameter space over which the sensitivity of the experiments to these parameters was determined.  The red star in the center of the plot indicates the position for the assumed true value of the parameters.  The mixing angle grid was chosen to be equally spaced in $\tan^{2}\theta$.  The grid in $\sin^{2}2\theta$ is non-linear, and is indicated in red below the grid points.}
   \label{fig:osc_par_grid}
\end{figure}

\noindent The likelihood function was evaluated in the anti-neutrino energy range $E_{\nu} \in \left[ 3.4, 10.0\right]$~MeV with 33 bins (bin width = 0.2~MeV).  Energy below 3.4~MeV was excluded to avoid irreducible background noise, which significantly diminishes the importance of the contribution of this energy region to the extraction of the oscillation parameters.  The quantity $\mu_{i}$ is the number of events expected for the current grid point in the oscillation parameter space, while $n_{i}$ is the randomly generated number of events for the central parameter pair.  The gamma function is just a convenient way to evaluate $n_{i}!$.  A total of 400 points in the parameter space was examined (Fig.~\ref{fig:osc_par_grid}).  At each grid point, data were generated randomly 1000 times and the log-likelihood function evaluated.  The average and RMS of the evaluations was taken to represent the log-likelihood value and uncertainty at that point.

The result for KamLAND is shown in Fig.~\ref{fig:osc_pars_marginalized_kl}, which is the marginalized log-likelihood (times 2) as a function of the oscillation parameters.  The 1$\sigma$ errors from the ensemble experiment is obtained by determining the distance from the minimum to the point where $2 \Delta {\cal L} = 1$; KamLAND's $\pm 1\sigma$ statistical errors are indicated by the vertical dashed blue lines.  The errors agree fairly well.  The $1\sigma$ error for $\Delta m^{2}_{12}$ for KamLAND according to the ensemble experiment is:

\begin{figure}[h] 
\begin{center}
   \includegraphics[width=10cm]{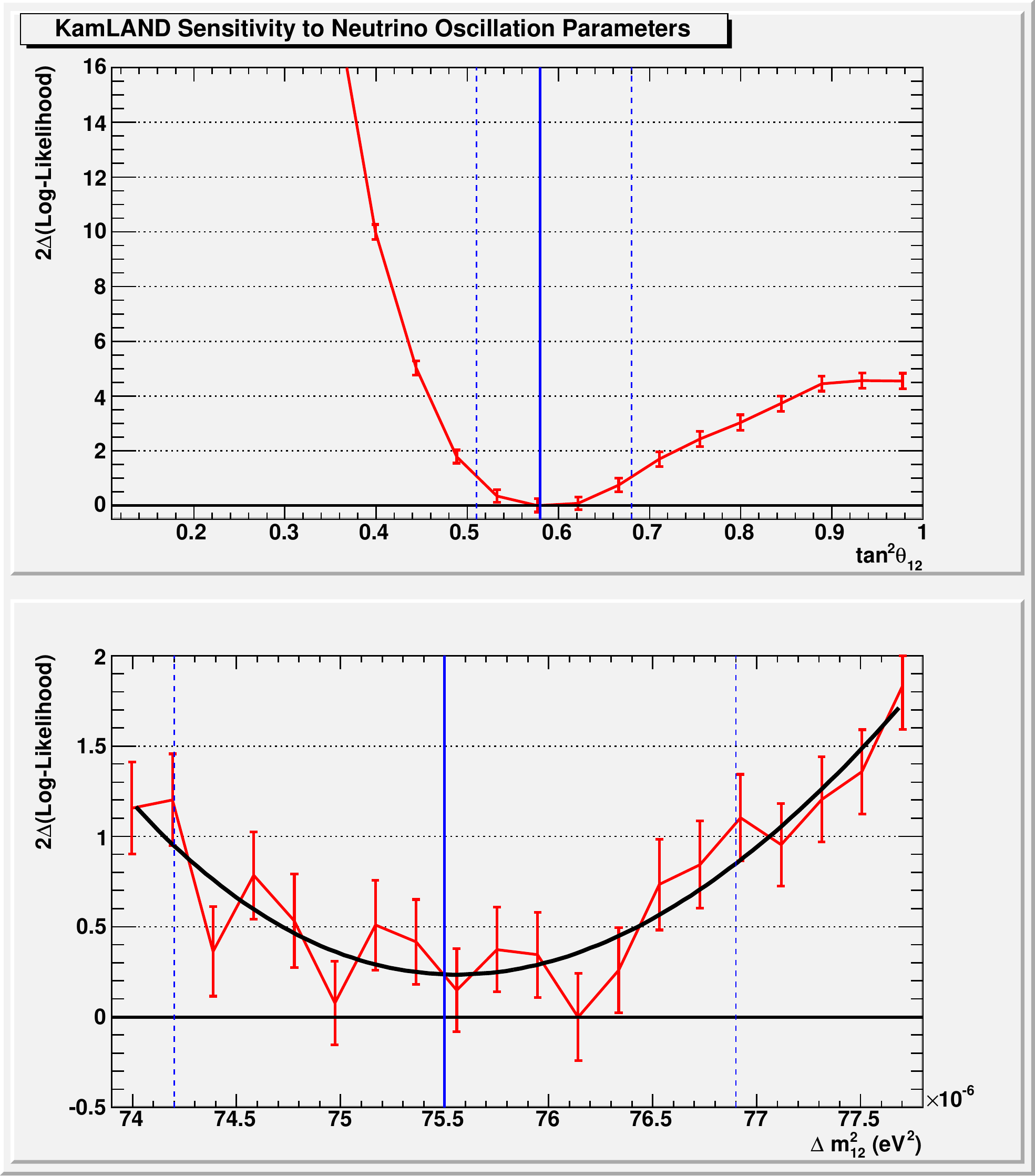} 
\end{center}
   \caption{The marginalized log-likelihood value (times 2) as a function of $\tan^{2}\theta_{12}$ (top) and $\Delta m^{2}_{12}$ (bottom).  The red points are the average log-likelihood value over 1000 determinations, while the error bars show the RMS spread.  In the bottom frame, the black curve through the points is the best-fit 3$^{rd}$ order polynomial.  The vertical blue line is the location of the minimum, while the dashed vertical blue lines show KamLAND's 1$\sigma$ statistical error.  The 1$\sigma$ error in this study is the distance from the minimum at which $2\Delta {\cal L} = 1$.}
   \label{fig:osc_pars_marginalized_kl}
\end{figure}

\begin{equation}
\label{eqn:dmsq_kl}
\Delta m^{2}_{12} = (7.55^{+0.19}_{-0.16}) \times 10^{-5} \; (\mbox{eV})^{2} \mbox{\hspace{3em} KamLAND}
\end{equation}

Fig.~\ref{fig:osc_pars_marginalized_sno+} shows the marginalized likelihood as a function of the oscillation parameters for SNO+.  As expected, SNO+ constrains the mixing angle much less because the statistics is much smaller, but the curvature of the likelihood as a function of $\Delta m^{2}_{12}$ is almost the same as for KamLAND.  The 1$\sigma$ limit for SNO+ is:

\begin{figure}[h] 
\begin{center}
   \includegraphics[width=10cm]{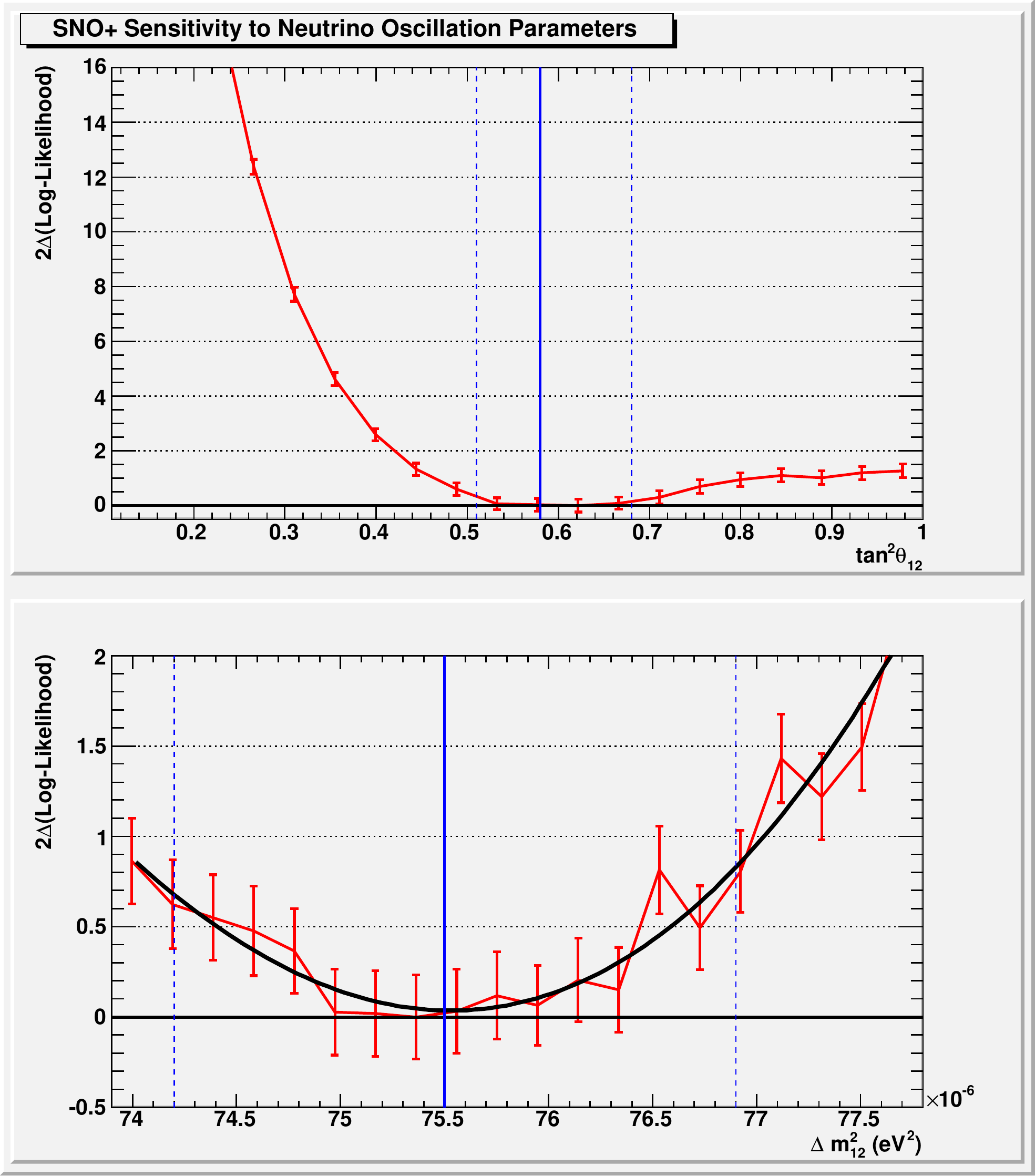} 
\end{center}
   \caption{Same as Fig.~\ref{fig:osc_pars_marginalized_kl}, but for SNO+.}
   \label{fig:osc_pars_marginalized_sno+}
\end{figure}

\begin{equation}
\label{eqn:dmsq_snoplus}
\Delta m^{2}_{12} = (7.55^{+0.16}_{-0.17}) \times 10^{-5} \; (\mbox{eV})^{2} \mbox{\hspace{3em} SNO+}
\end{equation}

In Eqn.~\ref{eqn:k_factor_eval}, the analysis based on the $K$-factor predicted that SNO+ would have $\Delta m^{2}$ sensitivity that is 20$\sim$30\% better than in KamLAND in spite of having only 1/5 the statistics.  Here, we find from Eqns.~\ref{eqn:dmsq_kl} and \ref{eqn:dmsq_snoplus} based on a rigorous likelihood analysis that the relative sensitivity is about the same (lower error) or about 20\% better (upper error).


\section{Conclusion}

Insofar as reactor anti-neutrino detection is concerned, the SNO+ detector is expected to perform similarly to KamLAND.  The main difference between the two is the distribution of nuclear reactors, which provide the detectors with anti-neutrinos.  KamLAND has a greater number of reactors within 300~km compared to SNO+, resulting in a flux that is about 5 times greater.  This gives KamLAND a natural statistical advantage of $\sqrt{5} \approx 2.2$ in determining the neutrino oscillation parameters for a given exposure to reactor anti-neutrinos.  SNO+, however, has an advantage based on the fact that relatively few reactors contribute to the bulk of detected events, and the few reactors that contribute to the bulk have baseline distances such that the oscillatory signature interfere constructively, resulting in the anti-neutrino spectrum shape varying rapidly as a function of $\Delta m^{2}_{12}$.  The advantage could be made quantitative in an approximate manner using the so-called $K$-factor technique.  This factor quantifies the degree of similarity of the shape of a function at two different parameter values: the more similar they are ({\it i.e.} the slower the shape changes with parameter value), the larger $K$ is.  Thus an experiment with a small $K$ is more sensitive to the parameter of interest compared to that with large $K$.  It was found that $K$ for SNO+ was 2.6$\sim$2.8 times smaller than in KamLAND, which more than compensates for the statistical disadvantage factor at SNO+ of 2.2.  Combining the effect of statistics and the shape, SNO+ is expected to outperform KamLAND in determining $\Delta m^{2}$ by about 20$\sim$30\%.  A more careful likelihood-based analysis confirmed this result: the sensitivity of SNO+ to $\Delta m^{2}_{12}$ was about the same or better by 20\% compared to that of KamLAND.

\appendix

\section{The Position and Thermal Power of Nuclear Reactors}
\label{app:reactors}

A list of nuclear reactor information was compiled based on information provided by the International Nuclear Safety Center (INSC) of the Argonne National Laboratory (ANL)~\cite{insc}.  As of 2006, there were 607 nuclear reactors in the list, although only a 478 of them were reported as operational.  The total rated thermal power of the reactors is 1.3~TW, while the total of the 478 operational ones is 1.1~TW.  A table summarizing this information is available at the following web site:

\begin{quote}
{\tt http://owl.phy.queensu.ca/$\sim$guillian/SNO+/reactor/2007-05-05/ \\
reactor\_list/Table.html}
\end{quote}

\noindent The power flux from these nuclear reactors is shown in Fig.~\ref{fig:power_flux}.  The power flux at a given location on Earth is given by:

\begin{figure}[h] 
\begin{center}
   \includegraphics[width=12cm]{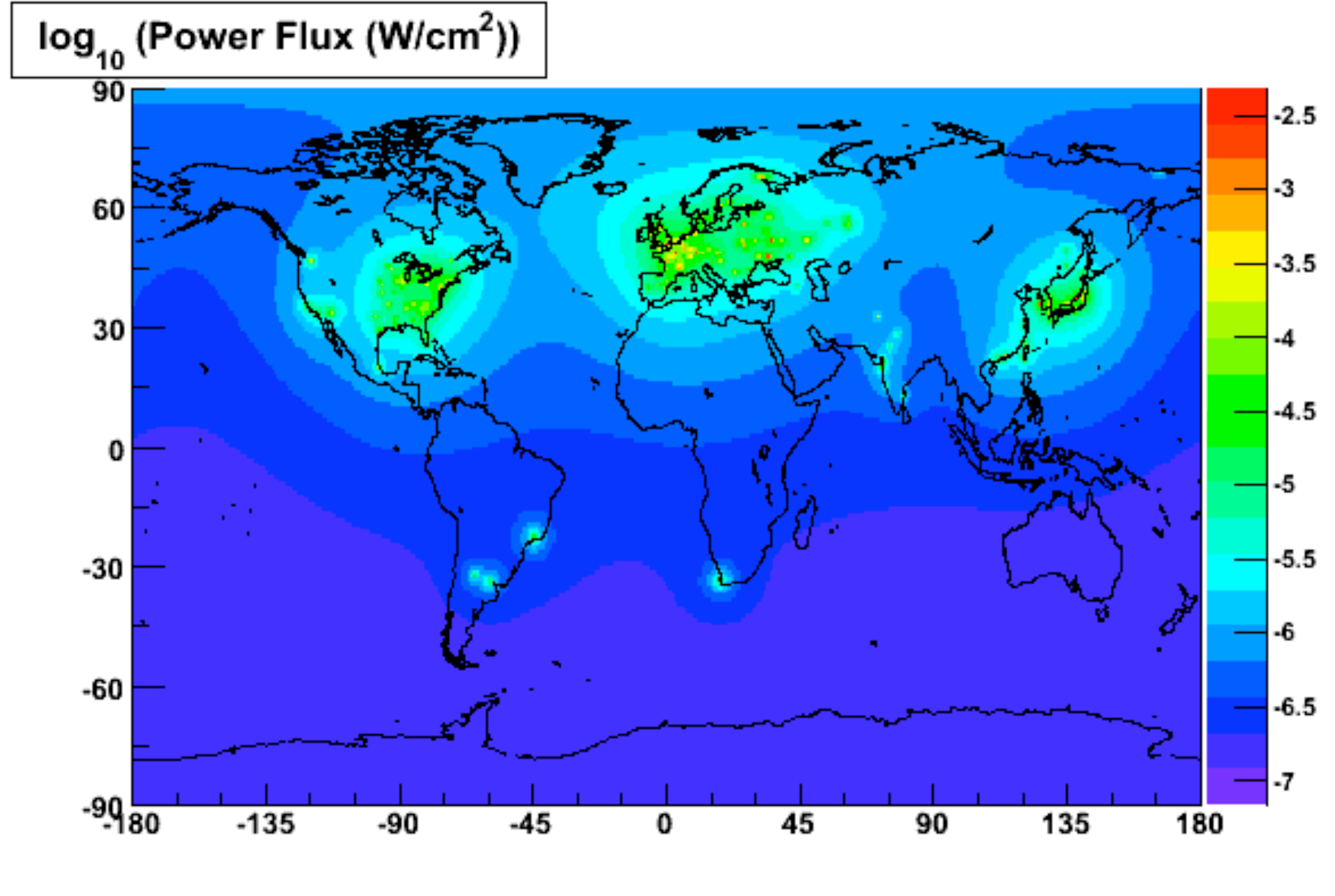} 
\end{center}
   \caption{The power flux from all operational nuclear reactors in the world.}
   \label{fig:power_flux}
\end{figure}

\begin{equation}
\sum_{i=1}^{N_{reactor}} \frac{P_{i}}{4 \pi \cdot D_{i}^{2}} \; ,
\end{equation}

\noindent where $P_{i}$ is the power of reactor number $i$, and $D_{i}$ is the distance between this reactor and the given point on Earth.

More information is available at the following web site:

\begin{quote}
{\tt http://owl.phy.queensu.ca/$\sim$guillian/SNO+/reactor/2007-05-05/ \\
index.html}
\end{quote}

\noindent In particular, the list of reactors is available in text file format in the following sites:

\begin{quote}
{\tt http://owl.phy.queensu.ca/$\sim$guillian/SNO+/reactor/2007-05-05/ \\
reactor\_list/reactor\_list\_2006-11-14.txt}
\end{quote}

\begin{quote}
{\tt http://owl.phy.queensu.ca/$\sim$guillian/SNO+/reactor/2007-05-05/ \\
reactor\_list/reactor\_list\_2006-11-14\_consolidated.txt}
\end{quote}

\noindent The first list gives the thermal power, operational status, and latitude and longitude of each reactor, while the latter gives information about only operational reactors.  Also note that the first one is a reactor-by-reactor list, while the latter is a site-by-site list, meaning that in sites with several reactors, the powers were combined.  When combining the powers, only operational powers were considered.  The total number of reactor sites is 208, whereas the total number of reactors is 607.

\bibliographystyle{plain}
\bibliography{reactor_sensitivity}

\end{document}